\documentclass[12pt]{article}
\usepackage{a4wide,epsfig}
\newcommand{\lmc}{l_{m_c}}

\begin{document}    

\title{\vskip-3cm{\baselineskip14pt
\centerline{\normalsize\hfill DESY 01--130}
\centerline{\normalsize\hfill TTP01--21}
\centerline{\normalsize\hfill hep-ph/0109084}
\centerline{\normalsize\hfill September 2001}
}
\vskip.7cm
Determination of $\alpha_s$ and heavy quark masses
from recent measurements of $R(s)$
}

\author{
{J.H. K\"uhn}$^{a}$
and
{M. Steinhauser}$^b$
  \\[3em]
  {\normalsize (a) Institut f\"ur Theoretische Teilchenphysik,}\\
  {\normalsize Universit\"at Karlsruhe, D-76128 Karlsruhe, Germany}
  \\[.5em]
  {\normalsize (b) II. Institut f\"ur Theoretische Physik,}\\ 
  {\normalsize Universit\"at Hamburg, D-22761 Hamburg, Germany}
}
\date{}
\maketitle

\begin{abstract}
\noindent
\vspace{.2cm}
In this paper we compare recent experimental data for the total cross
section $\sigma(e^+e^-\to\mbox{hadrons})$ with the up-to-date
theoretical prediction of perturbative QCD for those energies where
perturbation theory is reliable. The excellent agreement suggests the
determination of the strong coupling $\alpha_s$ from the measurements
in the continuum. The precise data from the charm threshold region,
when combined with the recent evaluation of moments with three loop
accurracy, lead to a direct determination of the short distance
$\overline{\rm MS}$ charm quark mass. Our result for the strong
coupling constant $\alpha_s^{(4)}(5~\mbox{GeV})=0.235^{+0.047}_{-0.047}$  
corresponds to $\alpha_s^{(5)}(M_Z)=0.124^{+0.011}_{-0.014}$, 
for the charmed quark mass we find $m_c(m_c)=1.304(27)$. Applying the
same approach to the bottom quark we obtain $m_b(m_b)=4.191(51)$~GeV. 
Whereas our result for $\alpha_s(M_Z)$ serves as a useful cross
check for other more precise determinations, our values for the charm
and bottom quark masses are more accurate than other recent analyses. 


\end{abstract}

\thispagestyle{empty}
\newpage
\setcounter{page}{1}

\renewcommand{\thefootnote}{\arabic{footnote}}
\setcounter{footnote}{0}


\section{Introduction}

The strong coupling constant and the quark masses are the basic input
parameters for the theory of strong interaction.
Quark masses are an essential input for the evaluation of weak decay
rates of heavy mesons and for quarkonium spectroscopy. 
The prediction of these masses is an important task for all variants
of Grand Unified Theories.
To obtain the values in a consistent way from different experimental
investigations is thus a must for current phenomenology.

During the past years new and more precise data for 
$\sigma(e^+e^-\to\mbox{hadrons})$ 
have become available in the low energy region between 2 and 10~GeV. 
At the same time increasingly precise calculations have been
performed in the framework of perturbative QCD (pQCD), 
both for the cross section as a
function of the center-of-mass energy $\sqrt{s}$,
including quark mass effects, and for its moments which
allow for a precise determination of the quark mass.
A fresh look at the determination of the strong coupling and the
quark masses based on these new developments is thus appropriate.

For the determination of $\alpha_s$ we shall use energy regions
sufficiently far away from the threshold where 
pQCD plus local duality is
expected to give reliable predictions. For the determination of the quark
masses we will use sum rules where the input depends heavily on the
threshold region with its rapidly varying cross section.

Sum rules as tool to determine the charm quark mass have been
suggested by the ITEP group long ago~\cite{NSVZ}. Subsequently these
methods have been developed further~\cite{Reinders:1985sr}
and frequently applied to the
bottom quark. Most of these later
analyses concentrated on using relatively
high moments which are less sensitive to the continuum contribution
and exhibit a very strong quark mass dependence.
However, 
this approach requires the proper treatment of the threshold, in part
the resummation of the higher order terms of the Coulombic binding
and a definition of the quark mass adopted
to this situation like the potential- or
$1S$-mass~\cite{Ben98,Hoang:1999nz}.
Recently improved experimental results in the charm
threshold region have become available~\cite{BES}, 
and at the same time the
moments have been evaluated up to order 
$\alpha_s^2$~\cite{CheKueSte96,CheKueSte97}. A fresh look at
the evaluation of the charm quark mass with the help of sum rules is
thus an obvious task.
We will concentrate on low moments, say up to the fourth one
and present the results for the fifth to the eighth moment just for
illustration.
This is a natural route to determine directly a short distance
mass, say $m_c(m_c)$ or even $m_c(3~\mbox{GeV})$ in the 
$\overline{\rm MS}$ scheme
as advocated in the original papers~\cite{NSVZ,ShiVaiZak79}.
In the present work we concentrate on the mass in the 
$\overline{\rm MS}$ scheme, $m_c(\mu)$, and adopt $\mu=3$~GeV as our
default value,
a scale characteristic for the present problem and sufficiently high
to ensure convergence of the perturbative series.
As already discussed in~\cite{NSVZ}, fixed order pQCD is adequate, if
only low moments are used in the analysis.

Recently the determination of the charm quark mass got quite some
attention.
In~\cite{Martin:2001dd} the pole mass $M_c$
has been determined from the comparison of the
direct determination of the hadronic contribution
to $\Delta\alpha(M_Z)$ with the determination using analytical
continuation. This leads to the range
$M_c=1.33-1.40$~GeV~\cite{Martin:2001dd}.
The pole mass has also been determined in~\cite{Eidemuller:2001rc}
using QCD sum rules in combination with nonrelativistic QCD.
Their result reads $M_c=1.70(13)$~GeV and is significantly higher than
the one of Ref.~\cite{Martin:2001dd}.
The $\overline{\rm MS}$ mass given in~\cite{Eidemuller:2001rc}
reads $m_c(m_c)=1.23(9)$~GeV.
On the basis of so-called Chauchy sum rules, recent experimental
data and analytical results for the three-loop photon polarization
function the $\overline{\rm MS}$ charm quark mass has been determined
in~\cite{Penarrocha:2001ig} with the result
$m_c(m_c)=1.37(9)$~GeV.
In~\cite{Narison:2001pu}
pseudo-scalar sum rules have been used to determine simultaneously the
decay constant $f_D$ and the charm quark mass.
For the latter the value $m_c(m_c)=1.10(4)$~GeV is given in the
abstract of~\cite{Narison:2001pu} which is significantly lower than
the other evaluations.

We also want to mention that there is a recent lattice evaluation of
the charm quark mass~\cite{Becirevic:2001yh} with the result
$m_c(m_c)=1.26(4)(12)$~GeV.
The first error corresponds to the statistical and the second
one to the systematical uncertainty.
Note that this result is derived in quenched QCD and the corresponding
uncertainty is not included.

In the remaining part of the Introduction we would like to fix the
notation and define the quantities we are dealing with in the
remainder of the paper.

It is convenient to normalize the radiatively corrected
hadronic cross section and to define the ratio
\begin{eqnarray}
  R(s) &=& \frac{\sigma(e^+e^-\to\mbox{hadrons})}{\sigma_{\rm pt}}
  \,,
\end{eqnarray}
where $\sigma_{\rm pt} = 4\pi\alpha^2/(3s)$.
As an inclusive quantity $R(s)$  is conveniently obtained via the optical
theorem from the imaginary part of the 
polarization function of two vector currents via
\begin{eqnarray}
  R(s) &=& 12\pi\,\mbox{Im}\left[ \Pi(q^2=s+i\epsilon) \right]
  \,,
\end{eqnarray}
where $\Pi(q^2)$ is defined through
\begin{eqnarray}
  \left(-q^2g_{\mu\nu}+q_\mu q_\nu\right)\,\Pi(q^2)
  &=&
  i\int {\rm d}x\,e^{iqx}\langle 0|Tj_\mu(x) j^{\dagger}_\nu(0)|0
  \rangle
  \,,
  \label{eq:pivadef}
\end{eqnarray}
with $j_\mu$ being the electromagnetic current.

The perturbative expansion of $R(s)$ can be written as
\begin{eqnarray}
  R(s) &=& \sum_Q
           R_Q^{(0)}
         + \frac{\alpha_s}{\pi} R_Q^{(1)}
         + \left(\frac{\alpha_s}{\pi}\right)^2 R_Q^{(2)}
         + \left(\frac{\alpha_s}{\pi}\right)^3 R_Q^{(3)}
         + \ldots
  \,,
  \label{eq:R}
\end{eqnarray}
where the summation is performed over all active quark flavours $Q$.
(We ignore the small singlet contribution at order $\alpha_s^3$.)
For a comprehensive compilation of the individual pieces 
we refer to~\cite{CheKueKwiPR,CheHoaKueSteTeu97} where also explicit
results are given. We want to stress that
the full quark mass dependence is available up to order
$\alpha_s^2$~\cite{CheKueSte96,CheKueSte97}.
In the case of $R_Q^{(3)}$ the first three terms in the
high-energy expansion are known~\cite{CheHarKue00}.

Our theoretical predictions are based on Eq.~(\ref{eq:R}) where the
up, down and strange quark masses are taken to be massless
and for the charm and bottom quark the respective pole masses are 
chosen as input.
If not stated otherwise we will use the following input values
for the evaluation of $R(s)$
\begin{eqnarray}
   \alpha_s^{(5)}(M_Z) &=&0.118\pm0.003
   \,,
   \nonumber\\
   M_c&=&(1.65\pm0.15)~\mbox{GeV}
   \,,
   \nonumber\\
   M_b&=&(4.75\pm0.20)~\mbox{GeV}
   \,,
   \label{eq:input}
\end{eqnarray}
which cover the full range of all currently accepted results.

At several places of our analysis the renormalization group functions
and the matching conditions for the masses and the strong coupling
are needed in order to get relations between different energy scales.
The corresponding calculations are performed using the package 
{\tt RunDec}~\cite{rundec}.

The outline of the paper is as follows:
in Section~\ref{sec:R} we compare the experimental data of $R(s)$ with
the theoretical prediction and determine $\alpha_s$ from the continuum 
data below $\sqrt{s}=3.73$~GeV and above $\sqrt{s}=4.8$~GeV.
The measurement in the charm threshold region is used for the
determination of the charm quark mass in Section~\ref{sec:charm}.
Similar considerations are used in Section~\ref{sec:bottom} in order
to obtain the bottom quark mass. Section~\ref{sec:con}
contains our conclusions.


\section{\label{sec:R}The continuum region}

\begin{figure}[t]
  \begin{center}
    \begin{tabular}{c}
      \leavevmode
      \epsfxsize=14cm
      \epsffile[40 250 550 580]{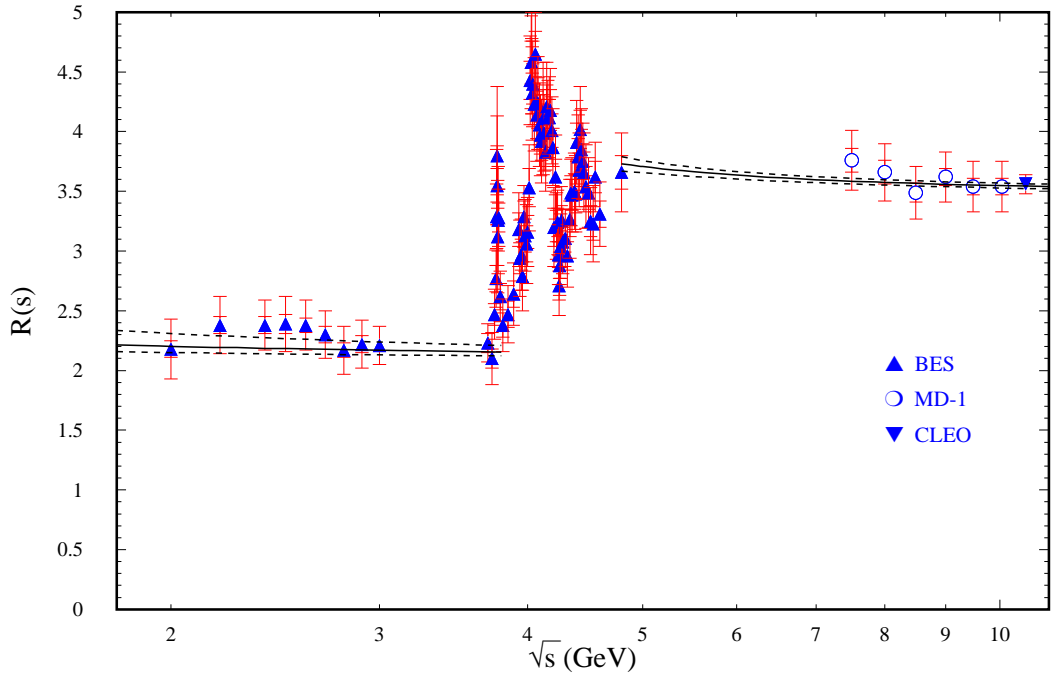}
    \end{tabular}
  \end{center}
  \vspace{-2.em}
  \caption{\label{fig:R}$R(s)$ in the energy range between $1.8$~GeV and
    $11.0$~GeV. The solid line corresponds to the theoretical
    predictions adopting our central values for the input parameters.
    The theoretical uncertainties are indicated by the dashed curves
    which are obtained from the variation of the input parameters as
    described in the text.
    The two error bars on the data points indicate the statistical
    (inner) and systematical (outer) uncertainty.
          }
\end{figure}

As stated in the Introduction, we distinguish
two energy regions: first,
the continuum region where pQCD and local duality
are expected to give reliable predictions for the hadronic cross
section and, second, the charm threshold region starting
from the $D$ meson threshold at $3.73$~GeV up to approximately
$5$~GeV, where the cross section exhibits rapid variations,
plus the $J/\Psi$ and $\Psi^\prime$ resonances.
The former will be mainly sensitive to the value of
$\alpha_s$. The latter will be used to evaluate moments and
to determine the charm quark mass.
For the present analysis the continuum region covers the 
BES data points from $2$~GeV up to $3.73$~GeV and the data from 
BES~\cite{BES}, MD-1~\cite{MD-1} and CLEO~\cite{CLEO} between $4.8$~GeV
and $10.52$~GeV\footnote{We limit this analysis to more recent results
  from BES~\cite{BES}, those from MD-1~\cite{MD-1} and from
  CLEO~\cite{CLEO} with systematic errors of typically $4.3$\%, $4$\%
  and $2$\%, respectively. Older measurements, in particular those from
  SPEAR and DORIS, are consistent with the new results. However, with their
  significantly larger errors they do not provide additional
  information.}.
As is evident from Fig.~\ref{fig:R} pQCD with
$\alpha_s^{(5)}(M_Z)=0.118$ provides an excellent description of all
recent results.

\begin{figure}[t]
  \begin{center}
    \begin{tabular}{cc}
      \leavevmode
      \epsfxsize=5cm
      \epsffile[203 302 540 507]{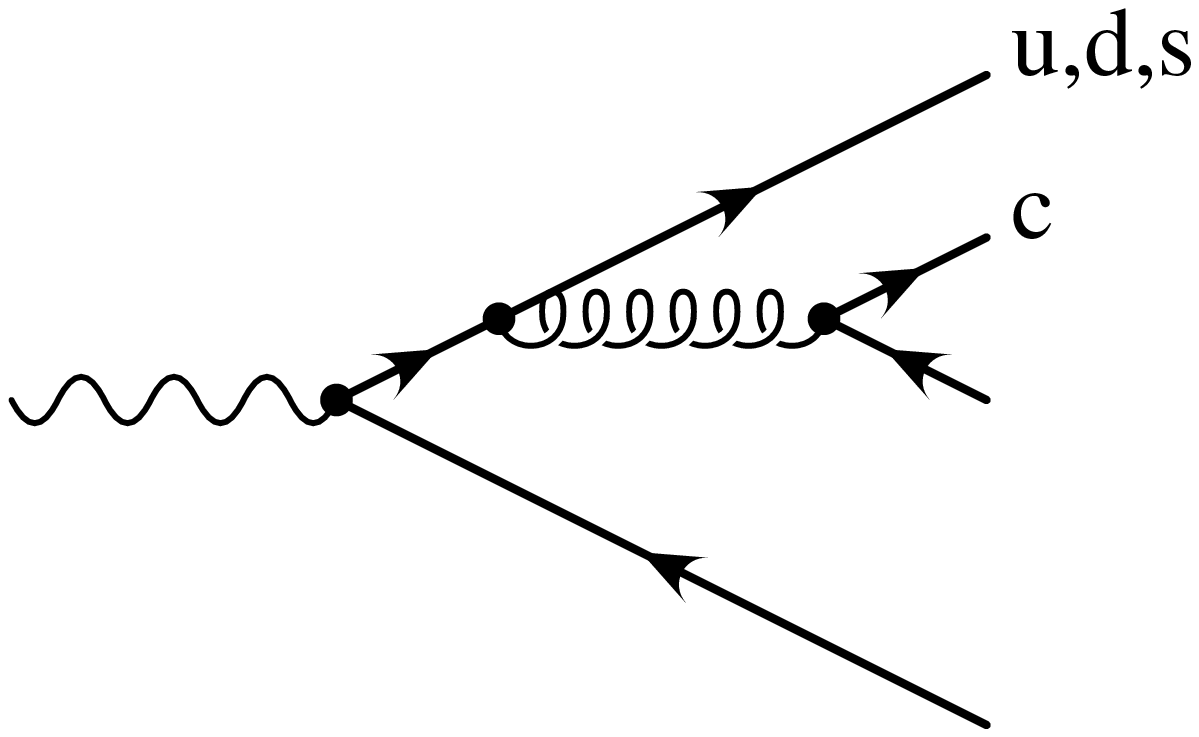}
      &
      \leavevmode
      \epsfxsize=5cm
      \epsffile[203 302 540 507]{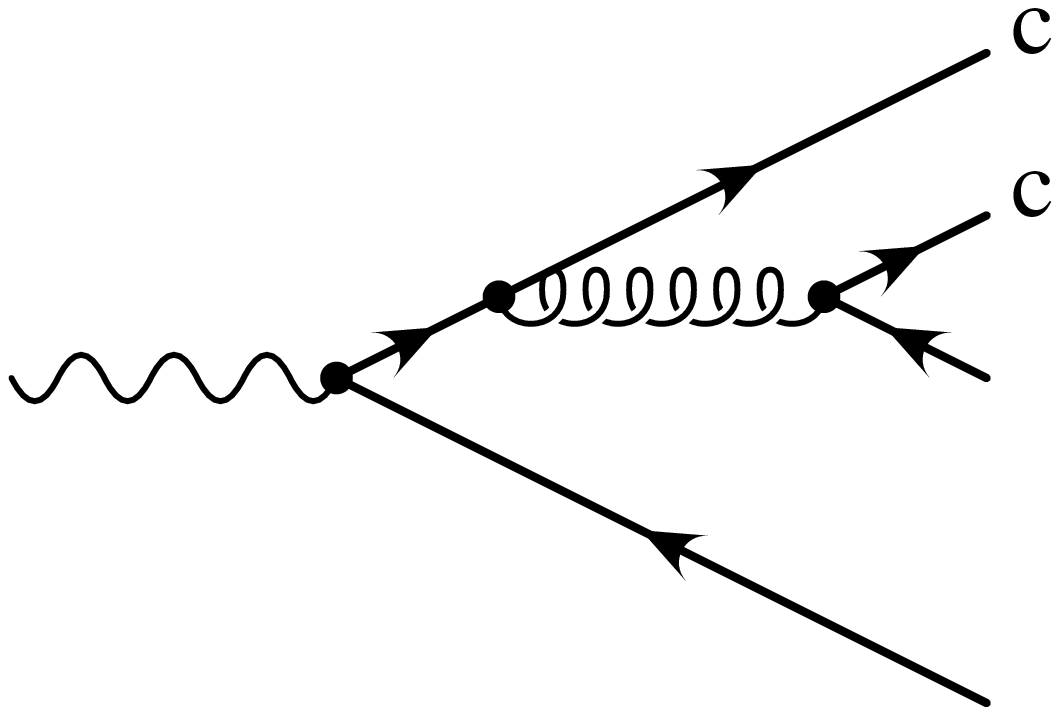}
      \\ (a) & (b)
    \end{tabular}
  \end{center}
  \caption{\label{fig:gacc}Feynman diagrams contributing to $R(s)$ at
    order $\alpha_s^2$. A secondary charm quark pair is 
    produced through gluon splitting.
          }
\end{figure}

Below $3.73$~GeV only $u$, $d$ and $s$ quarks are produced and the
${\cal O}(\alpha_s^3)$ approximation for massless quarks is adequate
for a description of $R$. The effective number of flavours is chosen
to be $n_f=3$ and virtual charm quark effects are taken into account
(for a compilation of the relevant formulae
see Ref.~\cite{CheKueKwiPR,CheHoaKueSteTeu97}).
Above charm threshold $u$, $d$ and $s$ quark production is calculated
as before, however, with $n_f=4$. Up to ${\cal O}(\alpha_s^2)$ the
prediction for the charm quark production incorporates the full
$M_c$ dependence. Starting from order $\alpha_s^2$ also the $M_c$
dependence of ``secondary'' charm production has to be taken into
account. This includes diagrams of the type in Fig.~\ref{fig:gacc}(a) 
as well as those from Fig.~\ref{fig:gacc}(b).
In addition we include ${\cal O}(\alpha_s^3)$ terms from the expansion
in $(M_c^2/s)^n$ with $n=0,1$ and $2$~\cite{CheKueKwiPR,CheHarKue00}.
Last not least contributions from virtual $c$ quarks
($\sqrt{s}\le3.73$~GeV) and $b$ quarks ($\sqrt{s}\le10.52$~GeV) have been
calculated which are suppressed $\sim (\alpha_s/\pi)^2 s/(4M^2)$ and
decouple for $s\ll4M^2$.
These are included in the fifth column of Tab.~\ref{tab:R}.
Pure QED final state radiation is tiny and taken into account for
completeness. 
(For a related analysis at $\sqrt{s} =10.5$~GeV 
see~\cite{Chetyrkin:1997tz}.)

\begin{table}[t]
\begin{center}
\begin{tabular}{r|r|r|r|r|r|r}
  $\sqrt{s}$ (GeV)& $n_f$           & $\alpha_s^{(n_f)}(\sqrt{s})$ & 
  $u$, $d$, $s$ & $c$, $b$        & $R^{\rm th}(s)$ & $R^{\rm exp}(s)$ 
  \\
  \hline
   2.000 & 3 &    0.297 &    2.202 &    0.002 &    2.203 &    2.180$\pm$   0.070\\
   3.730 & 3 &    0.226 &    2.154 &    0.002 &    2.156 &    2.100$\pm$   0.080\\
   4.800 & 4 &    0.215 &    2.146 &    1.607 &    3.753 &    3.660$\pm$   0.140\\
   8.900 & 4 &    0.180 &    2.122 &    1.440 &    3.562 &    3.579$\pm$   0.066\\
  10.520 & 4 &    0.172 &    2.117 &    1.427 &    3.544 &    3.560$\pm$   0.010\\
\end{tabular}
\caption{\label{tab:R}In the last two columns the experimental 
values for $R(s)$ are compared with 
the theoretical prediction, $R^{\rm th}(s)$, 
for a selected number of energies
where $\mu^2=s$ has been adopted (see text).
For completeness we list the number of active flavours, $n_f$, 
and the values of $\alpha_s^{(n_f)}(\sqrt{s})$.
}
\end{center}
\end{table}

The magnitude of the most important terms is shown for several
characteristic energies in Tab.~\ref{tab:R} and compared to the
experimental results. All numbers are given for the reference values
as specified in Eq.~(\ref{eq:input}).
The band in Fig.~\ref{fig:R} is obtained from the corresponding
errors and the theoretical uncertainty from the variation of the
renormalization scale between $\mu^2=s/4$ and $\mu^2=4s$
which have been added quadratically.
The excellent agreement between prediction and measurement 
suggests that one might determine $\alpha_s$ from this analysis. 

Assigning
a fully correlated systematic error of $4.3$\% to the BES 
data~\cite{Zhao_priv} we find
\begin{eqnarray}
  \alpha_s^{(3)}(3~\mbox{GeV}) &=&
  0.369^{+0.047}_{-0.046}{}^{+0.123}_{-0.130}
  \,,
  \label{eq:asBES_1}
\end{eqnarray}
from the BES data below $3.73$~GeV and 
\begin{eqnarray}
  \alpha_s^{(4)}(4.8~\mbox{GeV}) &=& 
  0.183^{+0.059}_{-0.064}{}^{+0.053}_{-0.057}
  \,,
  \label{eq:asBES_2}
\end{eqnarray}
from the point at $4.8$~GeV.
The uncertainties refer to the
uncorrelated (statistical) and correlated errors, respectively. 
The same analysis is applicable to the
MD-1 results. On the basis of the $R$ values listed for six energies 
in their table~4 and assuming fully correlated systematic errors
we find
\begin{eqnarray}
  \alpha_s^{(4)}(8.9~\mbox{GeV}) &=&
  0.201^{+0.026}_{-0.025}{}^{+0.129}_{-0.109}
  \,.
  \label{eq:asMD1_1}
\end{eqnarray}
From their average value
$R((8.9~\mbox{GeV})^2)=3.578\pm0.021\pm0.140$
as quoted in their summary
we obtain\footnote{This result disagrees from their value
  $\alpha_s^{(4)}(8.9~\mbox{GeV})=0.174\pm0.039$ 
  as a consequence of their
  inadequate treatment of the quark mass effects.}
\begin{eqnarray}
  \alpha_s^{(4)}(8.9~\mbox{GeV}) &=&
  0.193^{+0.017}_{-0.017}{}^{+0.127}_{-0.107}
  \,,
  \label{eq:asMD1_2}
\end{eqnarray}
consistent with Eq.~(\ref{eq:asMD1_1}) but with smaller errors.
From the CLEO value $R((10.52~\mbox{GeV})^2)=3.56\pm0.01\pm0.07$ we deduce
\begin{eqnarray}
  \alpha_s^{(4)}(10.52~\mbox{GeV}) &=&
  0.186^{+0.008}_{-0.008}{}^{+0.061}_{-0.057}
  \,.
  \label{eq:asCLEO}
\end{eqnarray}

We combine the results 
of Eqs.~(\ref{eq:asBES_1}),~(\ref{eq:asBES_2}),~(\ref{eq:asMD1_2})
and~(\ref{eq:asCLEO}) by assuming uncorrelated systematic
errors and evolving them to a common scale of $5$~GeV. Combining the
errors in quadrature we find
\begin{eqnarray}
  \alpha_s^{(4)}(5~\mbox{GeV}) &=&
  0.235^{+0.047}_{-0.047}
  \,.
\end{eqnarray}  
Evolving this value up to $M_Z$
\begin{eqnarray}
  \alpha_s^{(5)}(M_Z) &=&
  0.124^{+0.011}_{-0.014}
  \,,
\end{eqnarray}  
good agreement with other determinations~\cite{pdg00,Kniehl:2000cr}
is observed.


\section{\label{sec:charm}Charm 
  quark mass determination from the threshold region}

The first method we want to use for the determination of the charm quark
mass is based on the direct comparison of theoretical and experimental
moments of the charm quark contribution to the photon polarization
function as defined in Eq.~(\ref{eq:pivadef}).
In the limit of small momentum the latter can be cast 
into the form~\cite{CheKueSte97}
\begin{eqnarray}
  \Pi_c(q^2) &=& Q_c^2 \frac{3}{16\pi^2} \sum_{n\ge0}
                       \bar{C}_n z^n
  \,,
  \label{eq:pimom}
\end{eqnarray}
with $Q_c=2/3$ and 
$z=q^2/(4m_c^2)$ where $m_c=m_c(\mu)$ is the $\overline{\rm MS}$ 
charm quark mass at the scale $\mu$.
The perturbative series for the coefficients $\bar{C}_n$ 
up to $n=8$ is known
analytically~\cite{CheKueSte96,CheKueSte97} up to order $\alpha_s^2$.
The coefficients 
also depend on the charm quark mass through logarithms of the
form $\lmc\equiv\ln(m_c^2(\mu)/\mu^2)$ and can be written as
\begin{eqnarray}
  \bar{C}_n &=& \bar{C}_n^{(0)} 
              + \frac{\alpha_s(\mu)}{\pi}
                \left( \bar{C}_n^{(10)} + \bar{C}_n^{(11)}\lmc \right)
              + \left(\frac{\alpha_s(\mu)}{\pi}\right)^2
                \left( \bar{C}_n^{(20)} + \bar{C}_n^{(21)}\lmc
                       + \bar{C}_n^{(22)}\lmc^2 \right)
  \,.
  \nonumber\\
  \label{eq:cn}
\end{eqnarray}
In Tab.~\ref{tab:cn} the individual coefficients are given 
in numerical form.
They essentially constitute our theoretical input.
We define the moments
\begin{eqnarray}
  {\cal M}_n &\equiv& \frac{12\pi^2}{n!}
                      \left(\frac{{\rm d}}{{\rm d}q^2}\right)^n
                      \Pi_c(q^2)\Bigg|_{q^2=0}
  \,,
\end{eqnarray}
which leads to
\begin{eqnarray}
  {\cal M}_n^{\rm th} &=& 
  \frac{9}{4}Q_c^2
  \left(\frac{1}{4 m_c^2}\right)^n \bar{C}_n
  \,.
  \label{eq:Mth}
\end{eqnarray}

\renewcommand{\arraystretch}{1.1}
\begin{table}
\begin{center}
{\small
\begin{tabular}{r|rrrrrrrr}
  $n$ & 1 & 2 & 3 & 4 & 5 & 6 & 7 & 8 \\
  \hline
$\bar{C}_n^{(0)}$&$   1.0667$&$   0.4571$&$   0.2709$&$   0.1847$&$   0.1364$&$   0.1061$&$   0.0856$&$   0.0709$\\
$\bar{C}_n^{(10)}$&$   2.5547$&$   1.1096$&$   0.5194$&$   0.2031$&$   0.0106$&$  -0.1158$&$  -0.2033$&$  -0.2660$\\
$\bar{C}_n^{(11)}$&$   2.1333$&$   1.8286$&$   1.6254$&$   1.4776$&$   1.3640$&$   1.2730$&$   1.1982$&$   1.1351$\\
$\bar{C}_n^{(20)}$&$   2.4967$&$   2.7770$&$   1.6388$&$   0.7956$&$   0.2781$&$   0.0070$&$  -0.0860$&$  -0.0496$\\
$\bar{C}_n^{(21)}$&$   3.3130$&$   5.1489$&$   4.7207$&$   3.6440$&$   2.3385$&$   0.9553$&$  -0.4423$&$  -1.8261$\\
$\bar{C}_n^{(22)}$&$  -0.0889$&$   1.7524$&$   3.1831$&$   4.3713$&$
5.3990$&$   6.3121$&$   7.1390$&$   7.8984$\\
\end{tabular}
}
\caption{\label{tab:cn}Coefficients of the photon polarization
  function in the $\overline{\rm MS}$ scheme as defined in
  Eqs.~(\ref{eq:pimom}) and~(\ref{eq:cn}).
  $n_f=4$ has been adopted which is appropriate for the charm threshold.
  }
\end{center}
\end{table}
\renewcommand{\arraystretch}{1.0}

With the help of a dispersion relation we establish the connection
between the polarization function and the experimentally accessible 
cross section $R_c(s)$. In the $\overline{\rm MS}$ scheme
\begin{eqnarray}
  \Pi_c(q^2) &=& \frac{q^2}{12\pi^2}\int
  {\rm d}s\,\frac{R_c(s)}{s(s-q^2)}
  + Q_c^2 \frac{3}{16\pi^2} \bar{C}_0 
  \,,
  \label{eq:pidisp}
\end{eqnarray}
which allows to determine the experimental moments
\begin{eqnarray}
  {\cal M}_n^{\rm exp} &=& \int \frac{{\rm d}s}{s^{n+1}} R_c(s)
  \,.
  \label{eq:Mexp}
\end{eqnarray}
Note, that the last term in Eq.~(\ref{eq:pidisp}) 
which defines the renormalization scheme disappears after
taking derivatives with respect to $q^2$.
Equating Eqs.~(\ref{eq:Mth}) and~(\ref{eq:Mexp}) 
leads to an expression from which the charm quark mass can be obtained:
\begin{eqnarray}
  m_c(\mu) &=& \frac{1}{2} 
  \left(\frac{\bar{C}_n}{{\cal M}_n^{\rm exp}}\right)^{1/(2n)} 
  \,.
  \label{eq:mc1}
\end{eqnarray}

As a second method we consider the ratio of two successive moments
which leads to
\begin{eqnarray}
  m_c(\mu) &=& \frac{1}{2} 
  \sqrt{\frac{ {\cal M}_n^{\rm exp} }{ {\cal M}_{n+1}^{\rm exp} }
        \frac{ \bar{C}_{n+1} }{ \bar{C}_{n} }
       }
  \,.
  \label{eq:mc2}
\end{eqnarray}
Here the normalization uncertainty of the experimental data is largely
cancelled.
Both in Eqs.~(\ref{eq:mc1}) and~(\ref{eq:mc2}) one has to
remember the $m_c$ dependence of $\bar{C}_n$.

We have checked that the nonperturbative contribution from gluon
condensates~\cite{NSVZ,Broadhurst:1994qj}
can be neglected within the present accuracy.

Let us now turn to the extraction of the experimental values for the
three different 
contributions which enter the right-hand side of Eq.~(\ref{eq:Mexp}):
the $J/\Psi$ and $\Psi^\prime$
resonances (${\cal M}_n^{\rm exp,res}$), the charm threshold region between 
$2M_{D_0}\approx3.73$~GeV and $\sqrt{s_1}=4.8$~GeV as measured by the BES
experiment~\cite{BES}
(${\cal M}_n^{\rm exp,cc}$), and the continuum contribution above
$s_1$ (${\cal M}_n^{\rm cont}$).

The resonances are treated in the narrow width approximation
which corresponds to
\begin{eqnarray}
  R^{\rm res}(s) &=& \frac{9\pi M_R \Gamma_e}{\alpha^2}
                     \left(\frac{\alpha}{\alpha(s)}\right)^2
                     \delta(s-M_R^2)
  \,,
\end{eqnarray}
with $\alpha^{-1}=137.0359895$.
For both resonances we use $(\alpha/\alpha(s))^2\approx0.9562$
and
$M_{J/\Psi} = 3.09687(4)$~MeV,
$\Gamma_e^{J/\Psi} = 5.26(37)$~keV,
$M_{\Psi^\prime} = 3.68596(9)$~MeV
and 
$\Gamma_e^{\Psi^\prime} = 2.12(18)$~keV~\cite{pdg00}.

In the charm threshold region we have to identify the contribution from
the charm quark, i.e. we have to subtract the parts
arising from the light $u$, $d$ and $s$ quark from the data.
Technically this is done by determining a mean value for 
$R_{uds}\equiv R_u+R_d+R_s$ from the comparison of theoretical
predictions and the BES data between $2$~GeV and
$3.73$~GeV and using the theoretically predicted energy dependence to
extrapolate into the region between $3.73$~GeV and
$4.8$~GeV~\cite{KueSte98}.
This value is subtracted from the data before the integration is
performed.
Alternatively, one could adopt the massless prediction for 
$R(s)$ up to order $\alpha_s^3$ without taking into account the data
below $\sqrt{s}=3.73$~GeV. We checked that both approaches lead to the
same final result.

In the continuum region above $\sqrt{s}=4.8$~GeV 
there is only sparse and quite unprecise data. On the other hand 
pQCD provides reliable predictions for $R(s)$, which is essentially
due to the knowledge of the complete mass dependence up to order
$\alpha_s^2$~\cite{CheKueSte96}.
Thus in this region we will replace data by the theoretical prediction
for $R(s)$ as discussed in Section~\ref{sec:R}.

In Tab.~\ref{tab:Mexp} we present the results for the moments
separated according to the three different contributions
discussed above.
The error of the resonance contribution is due to the uncertainties
of the input parameters. In the case of the charm threshold
contribution the uncertainty is dominated by the correlated
normalization error of approximately 4.3\% of the BES data. 
In the continuum region we varied the input parameters as 
as stated in Eq.~(\ref{eq:input}) and 
the renormalization scale as $\mu=(3\pm1)$~GeV.
The errors of the three contributions are added quadratically.
It is illustrating to compare the composition
of the experimental error for the different moments. Generally speaking,
it is dominated by the resonance contribution, specifically by the
7\% and 9\% uncertainty in the leptonic widths of the $J/\Psi$ and
$\Psi^\prime$, respectively. 
For the moment with $n=1$ and to some extent the one
with $n=2$ the improvement in the cross section measurement due to BES
(from about 10 -- 20 \% systematic error down to 4.3\%) was important. The
parametric uncertainties (from $\alpha_s$ and $M_c$) and the residual
$\mu$-dependence which affect ${\cal M}_n^{\rm cont}$ are small. The higher
moments 
(in fact already for $n$ above two) are increasingly dominated by
the resonance contributions with their 7\% uncertainty.

\begin{table}[t]
\begin{center}
{
\begin{tabular}{l|lll|l}
$n$ & ${\cal M}_n^{\rm exp,res}$
& ${\cal M}_n^{\rm exp,cc}$
& ${\cal M}_n^{\rm cont}$
& ${\cal M}_n^{\rm exp}$
\\
 & $\times 10^{(n-1)}$
& $\times 10^{(n-1)}$
& $\times 10^{(n-1)}$
& $\times 10^{(n-1)}$
\\
\hline
$1$&$  0.1114(82)$ &$  0.0313(15)$ &$  0.0638(10)$ &$  0.2065(84)$ \\
$2$&$  0.1096(79)$ &$  0.0174(8)$ &$  0.0142(3)$ &$  0.1412(80)$ \\
$3$&$  0.1094(79)$ &$  0.0099(5)$ &$  0.0042(1)$ &$  0.1234(79)$ \\
$4$&$  0.1105(79)$ &$  0.0057(3)$ &$  0.0014(0)$ &$  0.1175(79)$ \\
$5$&$  0.1126(80)$ &$  0.0033(2)$ &$  0.0005(0)$ &$  0.1164(80)$ \\
$6$&$  0.1155(82)$ &$  0.0020(1)$ &$  0.0002(0)$ &$  0.1176(82)$ \\
$7$&$  0.1190(84)$ &$  0.0012(1)$ &$  0.0001(0)$ &$  0.1202(84)$ \\
$8$&$  0.1230(87)$ &$  0.0007(0)$ &$  0.0000(0)$ &$  0.1237(87)$ \\
\end{tabular}
}
\caption{\label{tab:Mexp}Experimental moments as defined in
  Eq.~(\ref{eq:Mexp}) separated according to the contributions from
  the resonances, the charm threshold region and the continuum region
  above $\sqrt{s}=4.8$~GeV.}
\end{center}
\end{table}

We use the results of Tab.~\ref{tab:Mexp} together with Eq.~(\ref{eq:mc1})
in order to obtain in a first step $m_c(3~\mbox{GeV})$.
Subsequently
the result is transformed to the scale-invariant mass
$m_c(m_c)$~\cite{rundec}
including the three-loop coefficients of the renormalization group
functions. Both results can be found in Tab.~\ref{tab:mc1}.
The starting value for 
$\alpha_s(3~\mbox{GeV}) = 0.254^{+0.015}_{-0.014}$
needed for this step is obtained from 
$\alpha_s(M_Z)= 0.118 \pm 0.003$ 
by using the renormalization group equations and the matching
conditions with four loop accuracy.

The errors listed in Tab.~\ref{tab:mc1} 
receive contributions from the uncertainties in the
experimental moments and the variation of $\alpha_s$
(cf. Eq.~(\ref{eq:input})) and $\mu=(3\pm1)$~GeV in the coefficients
$\bar{C}_n$ which are all added quadratically.
It is interesting to note that the uncertainty in $m_c(3~\mbox{GeV})$
induced by the experimental moments decreases from $0.028$ for $n=1$ to
$0.010$ for $n=4$ whereas the theoretical uncertainty 
from the renormalization scale increases from
$0.001$ to $0.064$, the one from $\alpha_s$ from $0.011$ to $0.019$.

The moment with $n=1$ is evidently least sensitive to nonperturbative
contributions from condensates, to the Coulombic higher order effects, the
variation of $\mu$ and the parametric $\alpha_s$ dependence.
Hence we adopt
\begin{eqnarray}
  m_c(m_c) &=& 1.304(27)~\mbox{GeV}
  \,.
  \label{eq:mcfinal}
\end{eqnarray}
as our final result. Using the two- and three-loop 
relation~\cite{GraBroGraSch90,CheSte99,MelRit99}
between the pole- and the $\overline{\rm MS}$-mass this corresponds to
\begin{eqnarray}
  M_c^{\rm(2-loop)} &=& 1.514(34)~\mbox{GeV}
  \,,
  \nonumber\\
  M_c^{\rm(3-loop)} &=& 1.691(35)~\mbox{GeV}
  \,.
\end{eqnarray}
Our result agrees within the uncertainties with the recent
determinations of
$m_c(m_c)$ in~\cite{Eidemuller:2001rc,Penarrocha:2001ig}
but is comparatively more precise.

\begin{table}[t]
\begin{center}
{
\begin{tabular}{l|llll}
\hline
$n$ & 1 & 2 & 3 & 4 \\
\hline
$m_c( 3~\mbox{GeV})$&$  1.027(30)$&$  0.994(37)$&$  0.961(59)$&$  0.997(67)$\\
$m_c(m_c)$&$  1.304(27)$&$  1.274(34)$&$  1.244(54)$&$  1.277(62)$\\
\hline
$n$ & 5 & 6 & 7 & 8 \\
\hline
$m_c( 3~\mbox{GeV})$&$  1.094(110)$&$  1.184(161)$&$  1.253(182)$&$  1.307(191)$\\
$m_c(m_c)$&$  1.366(100)$&$  1.447(146)$&$  1.510(165)$&$  1.558(172)$\\
\hline
\end{tabular}
}
\caption{\label{tab:mc1}Results for 
  $m_c(3~\mbox{GeV})$ and $m_c(m_c)$ 
  in GeV obtained from Eq.~(\ref{eq:mc1}).}
\end{center}
\end{table}

\begin{figure}[t]
  \begin{center}
    \begin{tabular}{c}
      \leavevmode
      \epsfxsize=14cm
      \epsffile[40 250 550 580]{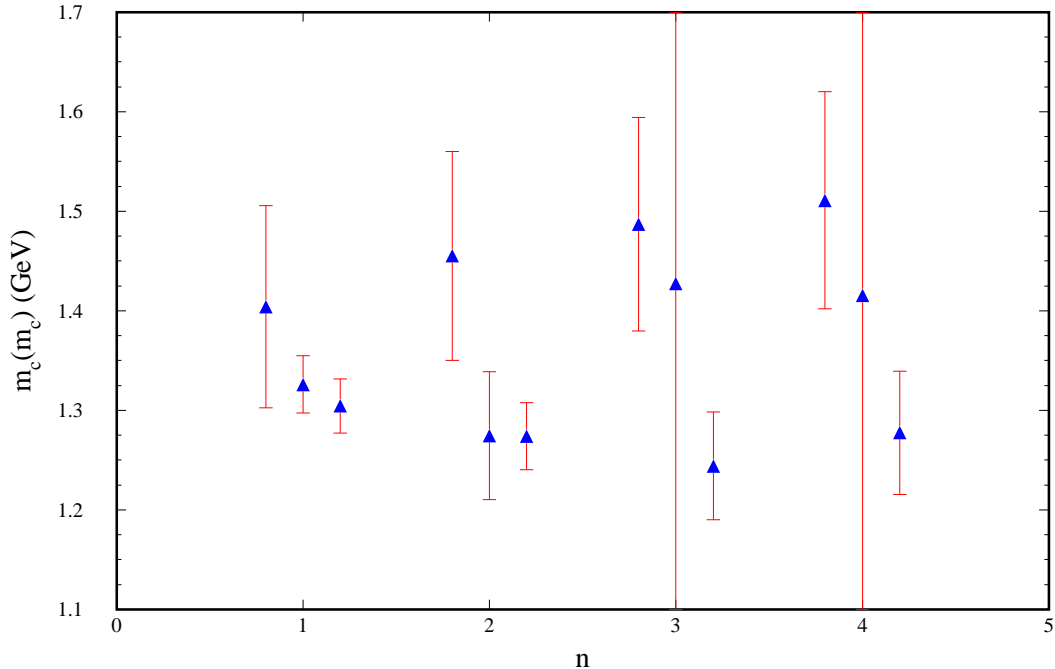}
    \end{tabular}
  \end{center}
  \vspace{-2.em}
  \caption{\label{fig:mom}$m_c(m_c)$ for $n=1,2,3$ and $4$.
  For each value of $n$ the results from left to right correspond
  the inclusion of terms of order $\alpha_s^0$, $\alpha_s^1$ and
  $\alpha_s^2$ in the coefficients $\bar{C}_n$ (cf. Eq.~(\ref{eq:cn})).
  Note, that for $n=3$ and $n=4$ the errors can not be determined
  with the help of Eq.~(\ref{eq:mc1})
  in those cases where only the two-loop corrections of order $\alpha_s$ are
  included into the coefficients $\bar{C}_n$ as the equation cannot be
  solved for $m_c(3~\mbox{GeV})$.
  }
\end{figure}

In Fig.~\ref{fig:mom} we compare the results for $m_c(m_c)$ (and its
$\mu$ dependence) based on the theory moments ($n=1,\ldots,4$)
evaluated up to ${\cal O}(\alpha_s^0)$, ${\cal O}(\alpha_s^1)$ and
${\cal O}(\alpha_s^2)$, respectively.
The improved stability with increasing order in
$\alpha_s$ is evident, and the preference for the first moment is clearly
visible.

As an alternative we also present the mass values derived from the ratio
of moments (Eq.~(\ref{eq:mc2})) in Tab.~\ref{tab:mc2}. 
The error is obtained from the 
quadratic combination of the uncertainties induced by
the resonances, the integration in the charm threshold region, the
continuum contribution and the independent 
variation of $\alpha_s$ and $\mu$
in the coefficients $\bar{C}_n$.
It is significantly larger than the
one based on 
the analysis of the moments. The numbers are therefore compatible
with Eq.~(\ref{eq:mcfinal}) but do not improve upon this result.

\begin{table}[t]
\begin{center}
{
\begin{tabular}{l|llll}
\hline
$n/(n+1)$ & $1/2$ & $2/3$ & $3/4$ & $4/5$ \\
\hline
$m_c( 3~\mbox{GeV})$&$  0.954(64)$&$  0.876(340)$&$  1.052(153)$&$  1.259(261)$\\
$m_c(m_c)$&$  1.238(59)$&$  1.166(318)$&$  1.327(140)$&$  1.515(236)$\\
\hline
$n/(n+1)$ & $5/6$ & $6/7$ & $7/8$ \\
\hline
$m_c( 3~\mbox{GeV})$&$  1.434(339)$&$  1.574(386)$&$  1.679(408)$\\
$m_c(m_c)$&$  1.671(304)$&$  1.795(344)$&$  1.886(361)$\\
\hline
\end{tabular}
}
\caption{\label{tab:mc2}Results for
  $m_c(3~\mbox{GeV})$ and $m_c(m_c)$
  in GeV obtained from Eq.~(\ref{eq:mc2}).}
\end{center}
\end{table}


\section{\label{sec:bottom}The bottom quark mass}

\begin{table}
\begin{center}
{\small
\begin{tabular}{r|rrrrrrrr}
  $n$ & 1 & 2 & 3 & 4 & 5 & 6 & 7 & 8 \\
  \hline
$\bar{C}_n^{(0)}$&$   1.0667$&$   0.4571$&$   0.2709$&$   0.1847$&$   0.1364$&$   0.1061$&$   0.0856$&$   0.0709$\\
$\bar{C}_n^{(10)}$&$   2.5547$&$   1.1096$&$   0.5194$&$   0.2031$&$   0.0106$&$  -0.1158$&$  -0.2033$&$  -0.2660$\\
$\bar{C}_n^{(11)}$&$   2.1333$&$   1.8286$&$   1.6254$&$   1.4776$&$   1.3640$&$   1.2730$&$   1.1982$&$   1.1351$\\
$\bar{C}_n^{(20)}$&$   3.1590$&$   3.2319$&$   2.0677$&$   1.2204$&$   0.7023$&$   0.4304$&$   0.3359$&$   0.3701$\\
$\bar{C}_n^{(21)}$&$   3.4425$&$   5.0798$&$   4.5815$&$   3.4726$&$   2.1508$&$   0.7592$&$  -0.6426$&$  -2.0281$\\
$\bar{C}_n^{(22)}$&$   0.0889$&$   1.9048$&$   3.3185$&$   4.4945$&$   5.5127$&$   6.4182$&$   7.2388$&$   7.9929$\\
\end{tabular}
}
\caption{\label{tab:cn5}Coefficients of the photon polarization
  function in the $\overline{\rm MS}$ scheme as defined in
  Eqs.~(\ref{eq:pimom}) and~(\ref{eq:cn}).
  $n_f=5$ has been adopted which is appropriate for the bottom
  threshold.
  }
\end{center}
\end{table}

The same approach is also applicable to the determination of
$m_b$. The coefficients $\bar{C}_n$ are listed
in Tab.~\ref{tab:cn5}. They determine the theoretical moments
through Eq.~(\ref{eq:Mth}) where $Q_c$ has to be replaced by $Q_b=-1/3$.
The experimental results for the moments are
listed in Tab.~\ref{tab:Mn5}. 
The contribution from the resonances include 
$\Upsilon(1S)$ up to $\Upsilon(6S)$ and are given by
${\cal M}_n^{\rm exp,res}$.
The treatment of the region between 
$\sqrt{s}=11.075$~GeV and $\sqrt{s}=11.2$~GeV
follows~\cite{KueSte98}. We assume a linear raise from zero to 
the pQCD prediction
$R_b((11.2~\mbox{GeV})^2)$ and the take the contribution itself as an
estimate for the error. Note that the contribution to the
moments, ${\cal M}_n^{\rm exp,lin}$ is  negligible small.
Above $11.2$~GeV we use the prediction from pQCD for $R_b(s)$ which
results in the moments ${\cal M}_n^{\rm cont}$.

\begin{table}[t]
\begin{center}
{
\begin{tabular}{l|lll|l}
$n$ & ${\cal M}_n^{\rm exp,res}$
& ${\cal M}_n^{\rm exp,lin}$
& ${\cal M}_n^{\rm cont}$
& ${\cal M}_n^{\rm exp}$
\\
  & $\times 10^{(2n+1)}$
& $\times 10^{(2n+1)}$
& $\times 10^{(2n+1)}$
& $\times 10^{(2n+1)}$
\\
\hline
$1$&$   1.508(114)$ &$   0.035(35)$ &$   2.913(21)$ &$   4.456(121)$ \\
$2$&$   1.546(109)$ &$   0.028(28)$ &$   1.182(12)$ &$   2.756(113)$ \\
$3$&$   1.600(106)$ &$   0.022(22)$ &$   0.634(8)$ &$   2.256(108)$ \\
$4$&$   1.671(104)$ &$   0.018(18)$ &$   0.381(5)$ &$   2.070(105)$ \\
$5$&$   1.761(103)$ &$   0.014(14)$ &$   0.244(3)$ &$   2.019(104)$ \\
$6$&$   1.870(104)$ &$   0.012(12)$ &$   0.162(2)$ &$   2.044(104)$ \\
$7$&$   1.998(105)$ &$   0.009(9)$ &$   0.111(2)$ &$   2.119(106)$ \\
$8$&$   2.149(108)$ &$   0.007(7)$ &$   0.078(1)$ &$   2.234(109)$ \\
\end{tabular}
}
\caption{\label{tab:Mn5}Moments for the bottom quark system.
     }
\end{center}
\end{table}

The uncertainties are completely analogous
to the charm quark case. The only difference concerns the 
renormalization scale for which we adopt $\mu=(10\pm5)$~GeV.

The results for $m_b(10~\mbox{GeV})$ and $m_b(m_b)$ are listed in
Tab.~\ref{tab:mb}.
A remarkable consistency and stability is
observed. For $n=1$ the error is dominated by the experimental input.
For $n=3$ we obtain $\pm 0.036$ from the experimental input,
$\pm 0.025$ from $\alpha_s$ and $\pm 0.020$ from the variation of $\mu$.

\begin{figure}[t]
  \begin{center}
    \begin{tabular}{c}
      \leavevmode
      \epsfxsize=14cm
      \epsffile[40 250 550 580]{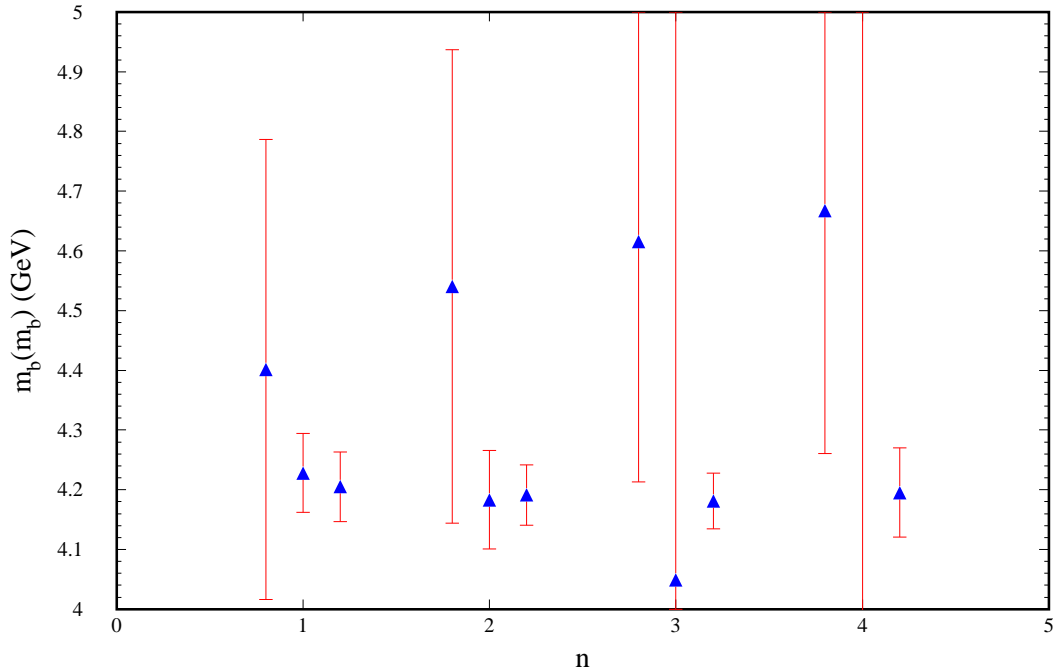}
    \end{tabular}
  \end{center}
  \vspace{-2.em}
  \caption{\label{fig:mbmb}$m_b(m_b)$ for $n=1,2,3$ and $4$.
  For each value of $n$ the results from left to right correspond
  the inclusion of terms of order $\alpha_s^0$, $\alpha_s^1$ and
  $\alpha_s^2$ in the coefficients $\bar{C}_n$ (cf. Eq.~(\ref{eq:cn})).
  Note, that the errors for $n=3$ and both the central value and the
  errors for $n=4$ can not be determined
  in those cases where only the two-loop corrections of order $\alpha_s$ are
  included into the coefficients $\bar{C}_n$ as the corresponding
  equation cannot be solved for $m_b(10~\mbox{GeV})$.
  }
\end{figure}

The sensitivity to the inclusion of higher orders is displayed in 
Fig.~\ref{fig:mbmb}.
Again a significant improvement of the stability of our prediction is
observed.
As our final result we adopt
\begin{eqnarray}
  m_b(m_b) &=& 4.191(51)~\mbox{GeV}
  \,,
  \label{eq:mbmb}
\end{eqnarray}
well consistent with evaluations based on the analysis of bottonium
and high spectral
moments~\cite{PenPiv1,MelYel,BenSig,Hoang:2000fm,Pineda:2001zq} 
(see also~\cite{pdg00})\footnote{Our result also agrees with the one
of Ref.~\cite{JamPic97}. However, 
in~\cite{JamPic97} the bottom quark mass has
been determined from relatively large moments ($7\le n\le15$), a highly
disputable treatment of the threshold has been employed and no stability for
small $n$ has been observed.}.
The $\overline{\rm MS}$ result of Eq.~(\ref{eq:mbmb}) corresponds to
a pole mass of~\cite{GraBroGraSch90,CheSte99,MelRit99}
\begin{eqnarray}
  M_b^{\rm(2-loop)} &=& 4.651(57)~\mbox{GeV}
  \,,
  \nonumber\\
  M_b^{\rm(3-loop)} &=& 4.819(57)~\mbox{GeV}
  \,,
  \label{eq:Mb}
\end{eqnarray}
using ${\cal O}(\alpha_s^2)$ and ${\cal O}(\alpha_s^3)$ accuracy,
respectively. 

\begin{table}[t]
\begin{center}
{
\begin{tabular}{l|llllllll}
\hline
$n$ & 1 & 2 & 3 & 4 \\
\hline
$m_b( 10~\mbox{GeV})$&$  3.665(60)$&$  3.651(52)$&$  3.641(48)$&$  3.655(77)$\\
$m_b(m_b)$&$  4.205(58)$&$  4.191(51)$&$  4.181(47)$&$  4.195(75)$\\
\hline
$n$ & 5 & 6 & 7 & 8 \\
\hline
$m_b( 10~\mbox{GeV})$&$  3.720(195)$&$  3.833(293)$&$  3.965(347)$&$  4.089(436)$\\
$m_b(m_b)$&$  4.258(189)$&$  4.367(285)$&$  4.494(335)$&$  4.614(420)$\\
\hline
\end{tabular}
}
\caption{\label{tab:mb}Results for
  $m_b(10~\mbox{GeV})$ and $m_b(m_b)$
  in GeV obtained from Eq.~(\ref{eq:mc1}) 
  as described in the text.}
\end{center}
\end{table}

\begin{table}[t]
\begin{center}
{
\begin{tabular}{l|llllllll}
\hline
$n/(n+1)$ & $1/2$ & $2/3$ & $3/4$ & $4/5$ \\
\hline
$m_b( 10~\mbox{GeV})$&$  3.636(52)$&$  3.620(51)$&$  3.694(240)$&$  3.915(523)$\\
$m_b(m_b)$&$  4.177(51)$&$  4.161(50)$&$  4.233(233)$&$  4.446(507)$\\
\hline
$n/(n+1)$ & $5/6$ & $6/7$ & $7/8$ \\
\hline
$m_b( 10~\mbox{GeV})$&$  4.218(717)$&$  4.523(875)$&$  4.786(1.113)$\\
$m_b(m_b)$&$  4.737(692)$&$  5.028(839)$&$  5.277(1.063)$\\
\end{tabular}
}
\caption{\label{tab:mb2}Results for 
  $m_b(10~\mbox{GeV})$ and $m_b(m_b)$
  in GeV obtained from Eq.~(\ref{eq:mc2})
  as described in the text.}
\end{center}
\end{table}


\section{\label{sec:con}Conclusions}

Recent experimental data for the total cross
section $\sigma(e^+e^-\to\mbox{hadrons})$ have been compared
with the up-to-date
theoretical prediction of perturbative QCD for those energies where
perturbation theory is reliable.
The excellent agreement justifies the
determination of the strong coupling $\alpha_s$ from the measurements
in the energy region between 2 and 3.73~GeV below the charm
threshold and the region between 4.8 and 10.52~GeV above charm and below the
bottom threshold. Our result 
$\alpha_s(5~\mbox{GeV})=0.235^{+0.047}_{-0.047}$ corresponds to
$\alpha_s(M_Z)=0.124^{+0.011}_{-0.014}$ and 
serves as a useful cross check in the intermediate
energy region but is less precise than those from $\tau$- or $Z$-boson
decays.

The direct determination of the short distance
$\overline{\rm MS}$ charm quark mass is performed
with the help of the precise data from the charm threshold
region and the three loop evaluation of moments in pQCD.
Using low moments the approach is insensitive to the Coulombic behaviour
of the cross section close to threshold and to nonperturbative
condensates. The results based on different moments are quite
consistent and the moment with $n=1$  exhibits the least sensitivity
towards the parametric dependence on $\alpha_s$ and the renormalization
scale. As our final result we obtain $m_c(m_c)=1.304(27)$~GeV.
The same approach when applied to the bottom quark
gives $m_b(m_b)=4.191(51)$~GeV.
These values are compatable with but more precise than other recent
analyses.


\section*{Acknowledgments}

We would like to thank A.A. Penin for carefully reading the manuscript.
We thank G. Corcella for drawing our attention to an error in
the running from $m_b(10~\mbox{GeV})$ to $m_b(m_b)$.
This work was supported in part by the {\it DFG-Forschergruppe
``Quantenfeldtheorie, Computeralgebra und Monte-Carlo-Simulation''} 
(contract FOR 264/2-1) and by SUN Microsystems through Academic
Equipment Grant No.~14WU0148.


\end{document}